\documentclass[aps,prmaterials,twocolumn,10pt,superscriptaddress]{revtex4-2}
\usepackage{amssymb,amsfonts,amsmath}
\usepackage{times}
\usepackage{graphicx}
\usepackage{color}
\usepackage{float}
\usepackage{cancel}
\newlength{\figurewidth}
\newlength{\pagewidth}
\begin{document}

\title{Structure, Stability and Superconductivity of N-doped Lutetium Hydrides at kbar Pressures}

\author{Katerina P. Hilleke}\thanks{These authors contributed equally}\affiliation{Department of Chemistry, State University of New York at Buffalo, Buffalo, NY 14260-3000, USA}
\author{Xiaoyu Wang}\thanks{These authors contributed equally}\affiliation{Department of Chemistry, State University of New York at Buffalo, Buffalo, NY 14260-3000, USA}
\author{Dongbao Luo}\affiliation{Department of Chemistry, State University of New York at Buffalo, Buffalo, NY 14260-3000, USA}
\author{Nisha Geng}\affiliation{Department of Chemistry, State University of New York at Buffalo, Buffalo, NY 14260-3000, USA}
\author{Busheng Wang}\affiliation{Department of Chemistry, State University of New York at Buffalo, Buffalo, NY 14260-3000, USA}
\author{Eva Zurek}\email{ezurek@buffalo.edu}\affiliation{Department of Chemistry, State University of New York at Buffalo, Buffalo, NY 14260-3000, USA}

\begin{abstract}
The structure of the material responsible for the room temperature and near ambient pressure superconductivity reported in an N-doped lutetium hydride [Nature, 615, 244 (2023)] has not been conclusively determined. Herein, density functional theory calculations are performed in an attempt to uncover what it might be. Guided by a range of strategies including crystal structure prediction and modifications of existing structure types, we present an array of Lu-N-H phases that are dynamically stable at experimentally relevant pressures. Although none of the structures found are thermodynamically stable, and none are expected to remain superconducting above $\sim$17~K at 10~kbar, a number of metallic compounds with \emph{fcc} Lu lattices -- as suggested by the experimental X-ray diffraction measurements of the majority phase -- are identified. The system whose calculated equation of states matches best with that measured for the majority phase is fluorite-type LuH$_2$, whose 10~kbar superconducting critical temperature was estimated to be 0.09~K using the Allen-Dynes modified McMillan equation.
\end{abstract}

\maketitle
\section{Introduction}

Heike Kamerlingh Onnes' 1911 discovery of mercury's entrance into a ``new...superconductive state'' at very low temperatures, where all electrical resistance vanished~\cite{Onnes:1911}, marked the beginning of a quest: could such a state be observed at room temperature? Ever since, scientists have sought this ``holy grail'', steadily breaking through barriers such as the boiling point of liquid nitrogen~\cite{Wu:1987}, 100~K~\cite{Sheng:1988}, then near 200~K~\cite{Drozdov:2015a,Israel:2022a,Minkov:2020a}. The latter breakthrough can be directly traced back to Ashcroft's proposal that hydrogen-rich alloys, metallized at conditions of extreme pressure, albeit less extreme than those required to metallize pure hydrogen, would be high-temperature phonon-mediated superconductors~\cite{Ashcroft:2004a}. It also marked a paradigm shift defined by a close synergy between theory and experiment, with computations either predicting the most promising superconducting phases or being instrumental in characterizing the synthesized compounds~\cite{Zurek:2021k,Zurek:2021n,Zurek:2018m}.

For the materials with the highest superconducting critical temperatures, $T_c$s,  that were found two things were true: they featured high hydrogen content and they required immense pressures -- approaching those found in the center of the Earth (350~GPa) -- for stability. One prominent class of these high-pressure high-temperature compounds are known as the ``superhydrides''. All of them are characterized by clathrate-like hydrogen-based lattices that encapsulate an electropositive metal atom, typically an alkaline or rare earth. Examples of compounds that have been both predicted and synthesized include CaH$_6$ ($T_c$ = 210-215~K, 160-172~GPa)~\cite{Ma:2022,Li:2022}, LaH$_{10}$ ($T_c$ = 260~K, 200~GPa) \cite{Somayazulu:2019,Drozdov:2019}, YH$_9$ ($T_c$ = 262~K, 182~GPa) \cite{Zurek:2020j}, YH$_6$ ($T_c$ = 224~K, 166~GPa) \cite{Troyan:2021a}, and mixed La/Y ternary hydrides~\cite{Semenok:2021c,Song:2021b} 

Clearly, the most prominent metal atoms in these phases are yttrium and lanthanum, with supporting roles played by calcium, scandium, and other rare earths. However, most of the heavier lanthanide hydrides are not expected to be as promising because of the suppressive influence of the \emph{f} electrons on superconductivity, with maximum $T_c$s decreasing rapidly once past La~\cite{SEMENOK2020100808,Peng:Sc-2017}. As a result, the hydrides of lutetium received relatively little attention despite the fact that the filled $4f$ shell of the metal is chemically unreactive rendering its electronic properties similar to those of Sc, Y and La... till now. 

An early theoretical study generated a Lu-H convex hull using known polyhydride structures, finding LuH$_4$, LuH$_6$, LuH$_9$, and LuH$_{10}$ as being thermodynamically stable at various pressures up to 400~GPa~\cite{Peng:Sc-2017}.  Another identified a unique $Immm$ structure for LuH$_8$ with an estimated $T_c$ of 81-86~K at 300~GPa, based on a distorted version of the backbone of the $Fm\bar{3}m$ LaH$_{10}$ phase~\cite{Sun:2020}. A theoretical comparison between the hydrides of the rare earth elements with filled vs.\ unfilled $f$-states -- Tm, Yb, and Lu, found LuH$_n$ (\emph{n}=4-8, 10) phases either on or very near the Lu-H convex hull at relatively low pressures (less than 200~GPa)~\cite{Song:2021}. Notably, LuH$_6$, with the same $Im\bar{3}m$ symmetry as CaH$_6$, had an estimated $T_c$ of 273~K (matching the melting point of ice) at 100~GPa. The filled $f$-shells of Lu and Yb were suggested to confer a strong degree of phonon softening, thereby resulting in a high electron-phonon coupling. Finally, a theoretical investigation of trends in superconductivity proposed high-pressure $Cc$ LuH$_7$ and $C222$ LuH$_{12}$ phases, with the latter predicted to undergo a superconducting transition below 6.7~K at 150~GPa~\cite{SEMENOK2020100808}. 

On the experimental side, a recent work reported the synthesis of a Lu hydride, suggested to be $Pm\bar{3}n$ Lu$_4$H$_{23}$, with a measured $T_c$ of 71~K at 218~GPa~\cite{arxivlu4h23}. This structure has previously been observed in experimental studies in the  La-H~\cite{Laniel:2022}, Ba-H~\cite{PenaAlvarez:2021}, and Eu-H~\cite{Semenok:2021} systems. 

Thus, with reported $T_c$s  of the superhydrides reaching temperatures not uncommon for a typical winter-day in upstate New York, the focus of research changed to predicting and synthesizing materials that could maintain high $T_c$s, but at lower pressures, with the ultimate goal of realizing superconductivity at ambient temperature and pressure. As the structures and superconducting properties of the binary hydrides had been exhaustively searched with no such candidate found, computations turned towards predicting ternary hydrides that remained dynamically stable to pressures below 100~GPa~\cite{DiCataldo:2021,Durajski:2021,Zhao:2022}, or boron-carbon analogues of the superhydrides that were stable at 1~atm~\cite{Zurek:2022m}.  

It was therefore quite exciting when a recent experimental manuscript reported superconductivity near room-temperature, $T_c$ = 294~K, at a very moderate pressure of 10~kbar (1~GPa) in a nitrogen-doped lutetium hydride phase~\cite{Dasenbrock:2023}. This pressure is low enough so that it becomes feasible to use pressure-quenching~\cite{Deng:2021a} to stabilize the material to ambient conditions, or to use careful strain engineering to achieve the desired superconductivity.  Unfortunately, though a variety of techniques including X-ray diffraction (XRD), energy-dispersive X-ray measurements, elemental analysis and Raman spectroscopy were used to characterize the superconducting material, its composition and structure could not be fully resolved~\cite{Dasenbrock:2023}.  

On the basis of the XRD and Raman analysis, the proposed room-temperature superconducting material (referred to as compound \textbf{A} by the authors) was indexed with space group $Fm\bar{3}m$, and both compound \textbf{A} and a minor product, which was dubbed compound \textbf{B}, were suggested to consist of an \emph{fcc} Lu network with additional N and H uptake~\cite{Dasenbrock:2023}. At pressures above $\sim$30~kbar, the superconducting compound \textbf{A} was found to undergo a pressure-induced transition to a non-superconducting structure involving a symmetry reduction of the Lu lattice to orthorhombic $Immm$ symmetry. The superconducting compound was also observed to undergo a sequence of color changes corresponding to structural transitions as pressure was applied, from blue to pink (marking transition to the high-$T_c$ superconductor), to red. 

Follow-up studies have, however, suggested that this color change is derived in fact from pure LuH$_2$~\cite{arxiv2303.06718,arxiv2303.08759}. Experiments reported no evidence for superconductivity down to 1.5~K in LuH$_2$~\cite{arxiv2303.06718}, or in LuH$_{2\pm x}$N$_y$ from ambient pressure to 6.3~GPa down to 10~K~\cite{arxiv2303.08759}. Moreover, DFT calculations~\cite{arxiv2303.06554} concluded that LuH$_2$ in the fluorite structure is the dominant phase of the parent nitrogen-doped superconductor, based on its computed thermodynamic and dynamic stability, optical properties and XRD pattern. A computational exploration of the Lu-N-H phase diagram found no ternary phases on the convex hull at pressures below 10~GPa, the binaries instead dominating, although a few ternary phases (Lu$_{20}$H$_2$N$_{17}$, Lu$_2$H$_2$N, LuH$_5$N$_2$, Lu$_3$H$_6$N, and Lu$_{10}$HN$_8$) were within 100~meV/atom of the hull. A number of the identified phases were found to be derived from either H vacancies or N-doping of LuH$_2$~\cite{arxiv2303.11683}. Another computational study did not find any thermodynamically stable Lu-N-H phases at 1~GPa and the highest $T_c$ computed for N-doped $Fm\bar{3}m$-LuH$_3$ did not exceed 30~K~\cite{arxiv2303.12575}.

Herein, we present a density functional theory (DFT) investigation of a series of structures in the Lu-N-H system that were either constructed via modification of known and theoretical prototype structures, via constrained and unconstrained crystal structure prediction (CSP) searches, or by a combination of these two methods. From the results of the unconstrained CSP runs we obtain a baseline against which to measure the enthalpies of constructed phases and to compare their properties. From constrained searches and artificially-constructed structures we begin to understand the motifs that contribute to dynamic stability at low pressures, and those which do not, allowing us to narrow the range of possible structures for further explorations into the Lu-N-H ternary system. The simulated X-ray diffraction patterns of the optimized phases and calculated equations of states are compared with available experimental data provided in Reference~\cite{Dasenbrock:2023}. The highest superconducting critical temperature we find --  17~K at 10~kbar -- was obtained for a CaF$_2$-type LuNH phase that was far from thermodynamic stability.

\section{Computational Details}
Precise geometry optimizations and electronic structure calculations were performed using DFT in conjunction with the Perdew-Burke-Ernzerhof (PBE) functional~\cite{Perdew:1996a}, as implemented in the Vienna \textit{ab initio} simulation package (VASP) \cite{kresse1996b,kresse1996efficient,kresse1999ultrasoft}.
The valence electrons of the hydrogen (H $1s^1$), nitrogen (N $2s^2 2p^3$), and lutetium (Lu $5p^{6} 5d^{1} 6s^{2}$) atoms were simulated using plane wave basis sets with a cutoff energy of 600~eV. The core electrons were treated with the projector augmented wave (PAW) method \cite{blochl1994projector}. Detailed tests of the inclusion of the $4f$ electrons on the properties of select structures, as well as the convergence of the plane wave basis were performed and representative results are provided in the Supporting Information. The reciprocal space was sampled using a $\Gamma$-centered Monkhorst-Pack mesh \cite{monkhorst1976special}, where the number of divisions along each reciprocal lattice vector was chosen such that the product of this number with the real-space lattice constant was 70~\r{A} for density of states calculations and 50~\r{A} for static calculations. To interrogate the dynamic stability of promising phases, phonon calculations were performed using the finite difference scheme, as implemented in the Phonopy software package \cite{Togo:2015,phonopy2}.

%The calculated results indicate that the $4f$ states of Lu mainly localize at the energy range from $-$4 to $-$6~eV and are significantly below the Fermi level. Thus, we do not expect them to participate in the superconductivity. Moreover, the POTCARs barely change the lattice in the case of $R3m$ Lu$_4$NH$_4$ (Volume: 373.79 vs.~374.36~\r{A}$^3$). To keep a good balance between accuracy and computational expense, Lu\_3 ($5p^6 6s^2 5d^1$) was selected to perform the remaining calculations (EOS, electronic properties, phonon dispersion, etc.) 

The electron-phonon coupling (EPC) calculations were performed using the Quantum Espresso (QE) package \cite{Giannozzi:2009,Giannozzi:2017} version 7.1 with the PBE functional. A plane wave basis set with a cutoff energy of 80 Ry was used, along with a charge density cutoff of 640 Ry for the valence electrons of hydrogen (H $1s^1$), nitrogen (N $2s^2 2p^3$), and lutetium (Lu $5s^2 5p^6 6s^2 5d^1$). The core electrons were treated with the PAW pseudopotentials generated using the PSLibrary package \cite{DalCorso:2014}. The $k$-point and $q$-point grids were selected to ensure the total electron-phonon-coupling (EPC) constant, $\lambda$, was converged to within 0.05 at the desired Gaussian broadening width for each structure, as summarized in the Supporting Information.

The superconducting critical temperature ($T\textsubscript{c}$) was
estimated using the Allen-Dynes modified McMillan equation \cite{Allen:1975}:
\begin{equation}
    T\textsubscript{c} =
\frac{\omega\textsubscript{ln}}{1.20}\exp\left[-\frac{1.04(1+\lambda)}{\lambda-\mu^*(1+0.62\lambda)}\right],
\end{equation}
in which the effective Coulomb potential, $\mu^*$, was set to 0.1, the
logarithmic average frequency $\omega\textsubscript{ln}$ was obtained by

\begin{equation}
\omega\textsubscript{ln}=\exp\left(\frac{2}{\lambda}\int\frac{d\omega}{\omega}\alpha^2F(\omega)\ln\omega\right),
\end{equation}
and the electron phonon coupling constant, $\lambda$, was evaluated by
\begin{equation}
    \lambda = \int d\omega \alpha^2F(\omega)/\omega.
\end{equation}
The Eliashberg spectral function, $\alpha^2F(\omega)$, was obtained from
the QE calculations, and was also used to numerically solve the Eliashberg
equations \cite{Eliashberg1960}.

The CSP searches were performed using the open-source evolutionary algorithm (EA)  \textsc{XtalOpt} \cite{Zurek:2011a,xtalopt-web,Zurek:2020i} version 12 \cite{Zurek:2018j}. The initial generation consisted of random symmetric structures created by the  \textsc{RandSpg} algorithm \cite{Zurek:2016h}. Duplicate structures were identified via the  \textsc{XtalComp} algorithm \cite{Zurek:2011i} and discarded from the breeding pool. Constrained \textsc{XtalOpt} searches were performed by determining the symmetry of the Lu atoms using Pymatgen~\cite{ong2013python} and only keeping those structures in the breeding pool that possessed an $Fm\bar{3}m$ symmetry Lu sublattice.  The parameters employed in the \textsc{XtalOpt} searches for the considered stoichiometries (number of formula units, pressures at which the EA searches were performed, and constraints employed) are provided in the Supporting Information.

\section{Results}

\subsection{Known Ambient Pressure Phases} 
Before we begin our theoretical investigation of novel Lu-N-H combinations that could be formed at mild pressures, let us review the structures and properties of the known LuH$_x$ and LuN phases. Unlike the high-pressure superhydrides, which bear little to no resemblance to the hydrides that are known at ambient conditions, the 1~atm LuN and LuH$_x$ phases may provide the key to the structure of Lu-N-H -- or at least very good starting points -- stemming from the relatively low pressures required to stabilize this ternary phase. 

\begin{figure}[!hbpt]
    \centering
    \includegraphics[width=\columnwidth]{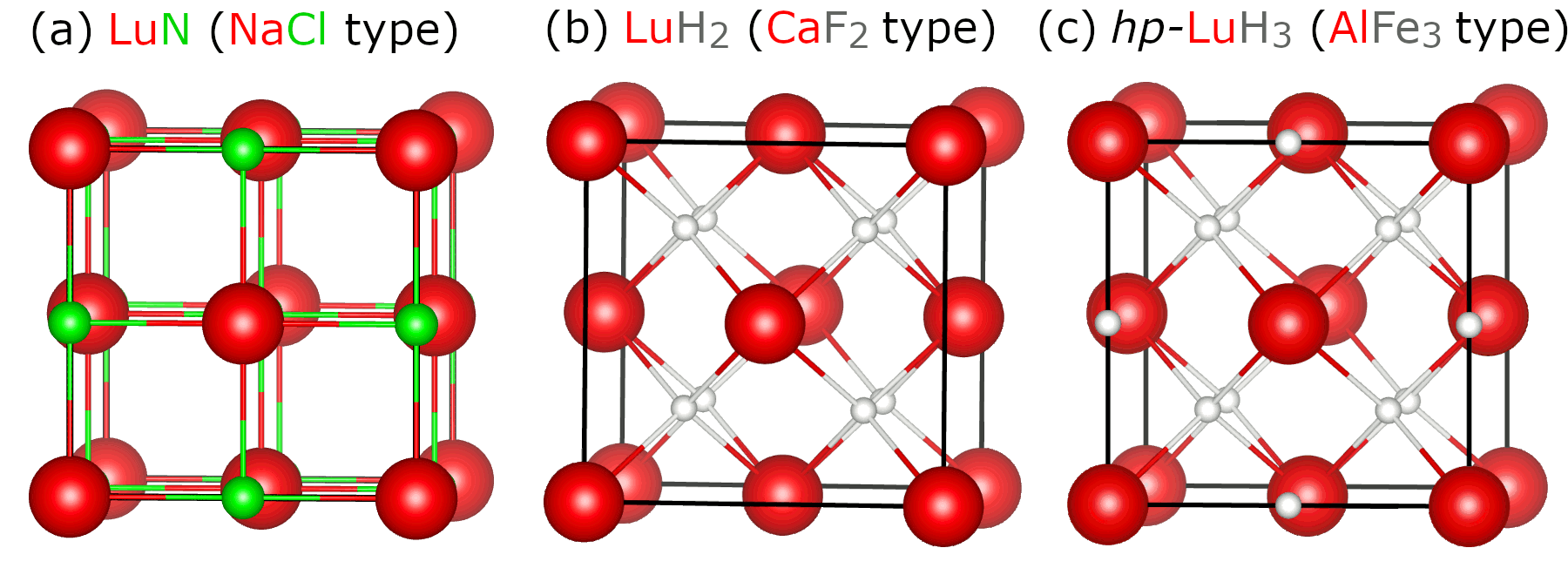}
    \caption{Prototype Lu-N and Lu-H structures with \emph{fcc} Lu lattices: (a)~NaCl-type LuN, (b) CaF$_2$-type LuH$_2$, and (c) a high-pressure (hp) phase of LuH$_3$.}
    \label{fig:prototypes}
\end{figure}

At ambient pressure, LuN assumes the rock-salt, or $B1$, structure (Figure \ref{fig:prototypes}a), with the Lu atoms in the \emph{fcc} configuration. A transition to the $B2$ or CsCl phase has been predicted near 250~GPa~\cite{Singh2015investigation}. Our PBE calculations, which likely underestimate the band gap, suggest semiconducting behavior at 1~atm with a gap of 0.23~eV.  In compounds, lutetium typically adopts the +3 oxidation state, and its hydrides can incorporate vacancies or extra hydrogen atoms that go into the interstitial regions~\cite{Libowitz:1972}. At 1~atm fluorite (CaF$_2$) type LuH$_x$ is adopted when $x=1.85-2.23$ (Figure \ref{fig:prototypes}(b)), usually resulting in a metallic phase. Increasing the hydrogen content to $x=2.78-3$ yields a hexagonal semiconducting phase~\cite{Libowitz:1972}. This $P\bar{3}c1$  LuH$_3$ transitions to a cubic phase at $\sim$10~GPa (the AlFe$_3$ or $D0_3$ structure type, Figure \ref{fig:prototypes}(c)), which can be stabilized at ambient via milling~\cite{Kataoka}. Recently, superconductivity in $Fm\bar{3}m$-LuH$_3$ was measured with a $T_c$ of 12.4~K at 122~GPa~\cite{Shao:2021a}.

To validate the computational settings used, we compared the lattice constants of the known phases where the Lu atoms are found in the $fcc$ arrangement: rock salt LuN (4.760~\r{A} \cite{Pebler:1962}) and fluorite type LuH$_2$ (5.033~\r{A} \cite{Olcese:1979}) with those of the optimized structures. The DFT lattice constants differed by only 0.17\% and 0.28\% from experiment, further supporting the choice of our computational parameters.  These known ambient-pressure nitrides and hydrides of lutetium provide a basis that could be used to build models of the high-$T_c$ superconducting phase reported in Ref.\ \cite{Dasenbrock:2023}. In fact, the similarity of the 1~atm lattice parameters of phase \textbf{A} (5.0289(4)~\r{A}) and the (presumably non-superconducting) compound \textbf{B} (4.7529~\r{A}) with the known dihydride and nitride of lutetium, respectively, coupled with a  comparison of the DFT-optimized unit cell parameters of several hypothetical and selected partially-doped versions of the known compounds were used to assign possible compositions~\cite{Dasenbrock:2023}. Phase \textbf{A} was tentatively assigned as LuH$_{3-\delta}$N$_{\epsilon}$, with partial N substitution onto H sites in the cubic (high-pressure)  LuH$_3$, and phase \textbf{B} as LuN$_{1-\delta}$H$_{\epsilon}$, an H-doped variant of rock-salt LuN~\cite{Dasenbrock:2023}.  On the other hand, a recent theoretical manuscript proposed that CaF$_2$-type LuH$_2$ is the parent structure of the superconducting phase, and compound \textbf{B} could be the rock-salt LuH structure, which is dynamically stable at 0~GPa~\cite{arxiv2303.06554}.

\subsection{Newly Predicted Phases}
The structures investigated herein were generated using a variety of procedures including \emph{ab initio} CSP techniques, as well as modification of known phases and compounds predicted using CSP.  The advantage of CSP searches is that they can, freed from structural preconceptions, locate the low-lying configurations in a potential energy surface, whose complexity here is heightened by the inclusion of three elements. Such searches can be unconstrained, purely hunting down the lowest-enthalpy configurations given a certain stoichiometry. Constraining a search to structures containing a particular motif will narrow down the possible results, but could also miss out on even lower-enthalpy alternatives that do not align with the constraints.  

To that end, a combination of both unconstrained and constrained CSP searches were carried out for the Lu-N-H system using the \textsc{XtalOpt} EA. From the former we can learn about the structural motifs that yield the most stability, and comparison with the latter informs us of the enthalpic cost associated with a specific structural feature. In addition, various structures were made ``by hand'' via modification of known prototypes or CSP generated structures that possess an \emph{fcc} Lu lattice. As we will soon see, a large structural variety is present amongst the dynamically stable phases that we found, highlighting the difficulties inherent in the computational prediction of metastable phases that could potentially be synthesized.\\

\noindent\textbf{Semiconductors:}

\begin{figure}[!hbpt]
    \centering
    \includegraphics[width=\columnwidth]{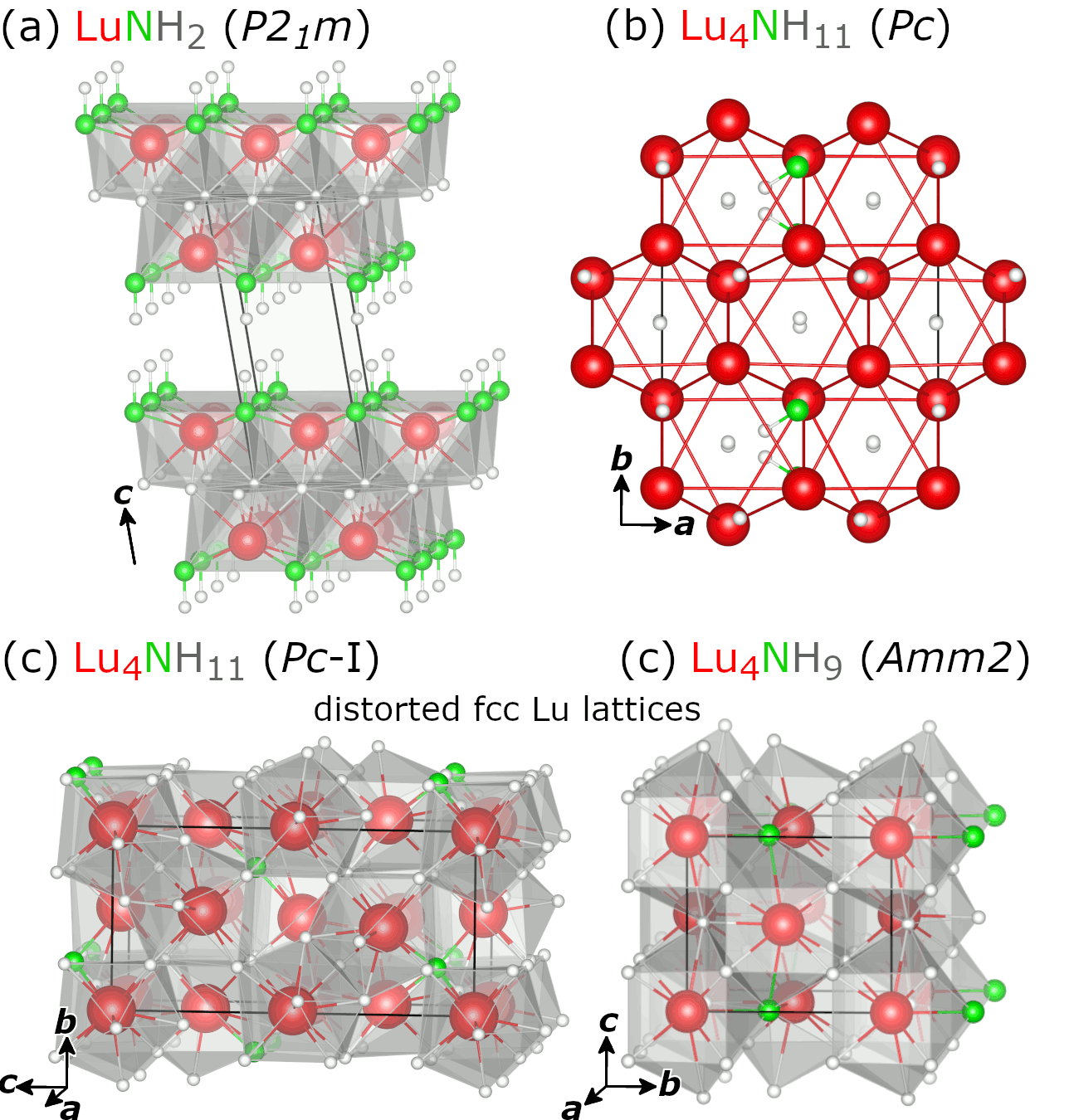}
    \caption{Semiconducting Lu-N-H phases found using unconstrained evolutionary crystal structure searches and prototype modification (Lu$_4$NH$_9$).}
    \label{fig:structures1}
\end{figure}

Unconstrained \textsc{XtalOpt} searches for the lowest-enthalpy structures were performed for the Lu$_3$NH$_{11}$ and Lu$_4$NH$_{10, 11}$ compositions at both 0 and 3~GPa, as well as for Lu$_4$NH$_6$ and LuNH$_2$ at~0 GPa. These EA runs located a number of structurally diverse semiconducting phases with PBE band gaps that ranged from 1.1-2.1~eV; some are shown in Figure~\ref{fig:structures1}. A few of the predicted structures, including $P2_1m$ LuNH$_2$, and two Lu$_3$NH$_{11}$ phases -- one with $P1$ symmetry at 0~GPa and one with $Cm$ symmetry at 3~GPa -- possessed large empty regions.  $P2_1m$ LuNH$_2$ (Figure \ref{fig:structures1}(a)) is, in fact, a fully 2D compound. At 0~GPa $Pc$ Lu$_4$NH$_{11}$ (Figure~\ref{fig:structures1}(b)) was also identified; it consists of layers of trigonal nets of Lu with H atoms in the resulting hexagonal channels, while the N atoms are arranged in zigzag chains oriented along the \emph{c}-axis that weave into the the Lu network (into the plane of the page). 

Two of the structures from unconstrained searches -- $P1$ Lu$_4$NH$_{10}$ (at 0~GPa) and a second $Pc$ Lu$_4$NH$_{11}$ structure (at 3~GPa; Figure~\ref{fig:structures1}(c)) -- possessed Lu sublattices in slightly distorted \emph{fcc} arrangements. In $P1$ Lu$_4$NH$_{10}$, the N atoms go into some of the sites octahedrally coordinated by Lu, while some H atoms go into the tetrahedral interstices and the rest are scattered across the unit cell, resulting in the very low symmetry. For $Pc$-I Lu$_4$NH$_{11}$ (Figure \ref{fig:structures1}(c)), the N atoms go instead into the tetrahedral interstices and the hydrogen atoms take the octahedral and most of the remaining tetrahedral interstices. The \emph{fcc} Lu lattice is also preserved in a semiconducting $Amm2$ compound with Lu$_4$NH$_9$ stoichiometry (Figure \ref{fig:structures1}(d)), which was produced not by CSP but by modifying the geometry of the high-pressure AlFe$_3$-type LuH$_3$ compound. Here, H again partially occupies both tetrahedral and octahedral interstices in \emph{fcc} Lu, leaving 1/4 of the tetrahedral interstices empty and with 1/4 of the H atoms filling octahedral interstices being replaced by N.

From these results, it is clear that a variety of geometric motifs can be found in the low-enthalpy Lu-N-H compounds, highlighting both the difficulty of honing in on a single structure and the utility of guidance from experimental data. The unit cell volumes of several of the systems identified via unconstrained CSP searches were too large for them to stay as candidates for the putative superconducting phase. Importantly, because all of the aforementioned compounds were semiconducting it is impossible for any of them to be superconductors. Our search continues, with inspiration taken from known experimental phases or CSP searches guided via constraints towards desired structural features -- or both. \\

\noindent\textbf{Structures from Prototype Modification:}

The relatively low pressures needed to stabilize the putative room-temperature superconducting phase highlight the importance of -- and inspiration to be gleaned from -- examining the ambient- and low-pressure compounds formed between Lu and either N or H. Notably, within most of these, the Lu atoms adopt the \emph{fcc} arrangement that has been suggested for the superconducting phase. 

\begin{figure}[!hbpt]
    \centering
    \includegraphics[width=\columnwidth]{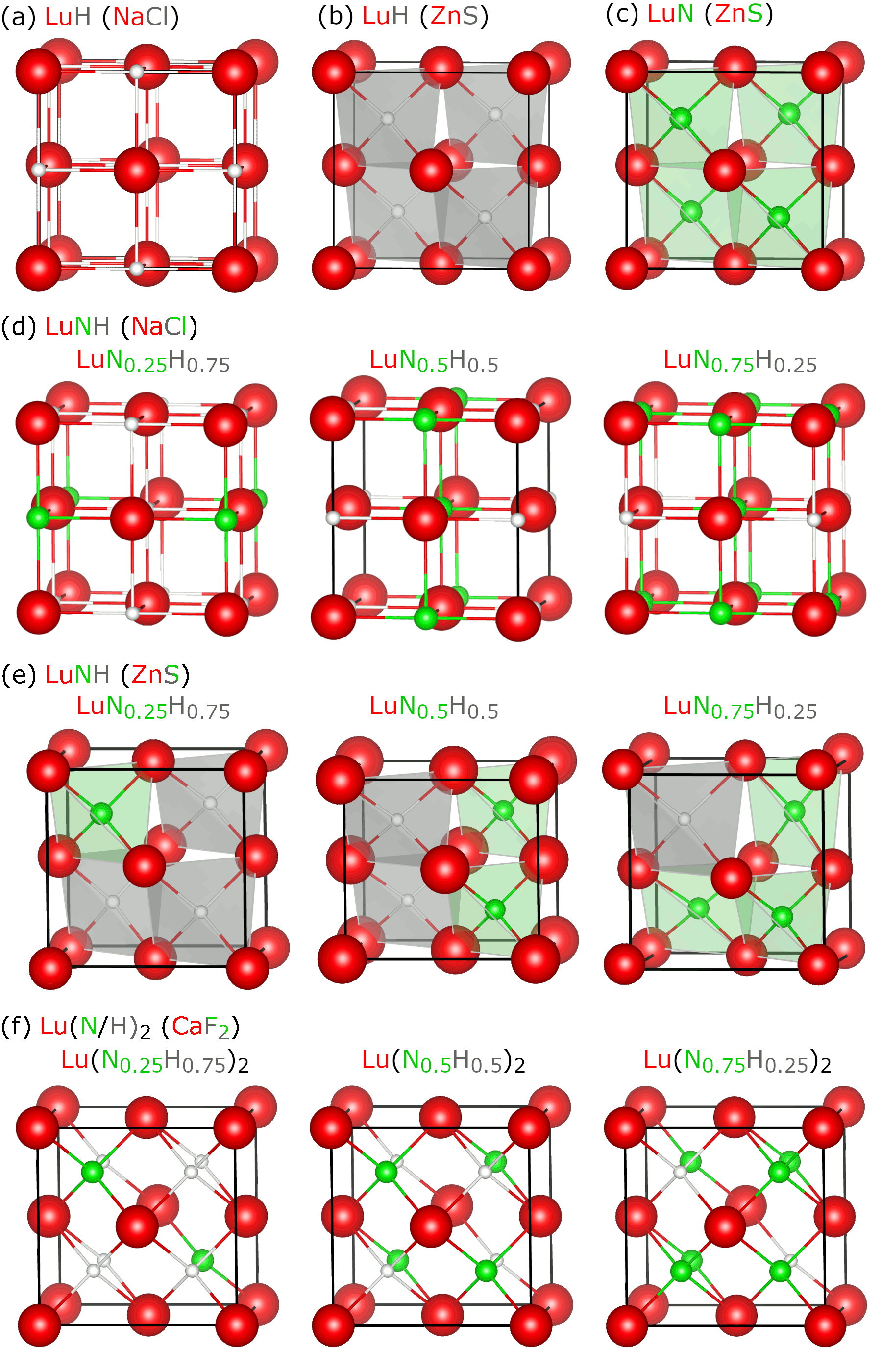}
    \caption{Illustrations of hypothetical (a) rock-salt ($B1$) LuH, and zinc-blende ($B3$) (b) LuH and (c) LuN phases. (d) Rock-salt and (e) zinc-blende LuN$_x$H$_{(1-x)}$, and (f) fluorite ($C1$) Lu(N$_x$H$_{(1-x)}$)$_2$ solid solution models that were considered.}
    \label{fig:structures2}
\end{figure}

In addition to the ambient pressure $B1$ mononitride, LuN (Figure \ref{fig:prototypes}(a)), we considered a hypothetical rock-salt monohydride, LuH (Figure \ref{fig:structures2}(a)), and hypothetical zinc-blende (or $B3$) LuN and LuH phases (Figure \ref{fig:structures2}(b,c)). To explore the potential of a solid solution between the two rock-salt phases  calculations were carried out on the unit cells shown in  Figure \ref{fig:structures2}(d). From these, only LuN$_{0.25}$H$_{0.75}$ and LuN$_{0.5}$H$_{0.5}$ were dynamically stable at 0~GPa. Similarly, solid solutions of zinc-blende LuN and LuH were optimized (Figure \ref{fig:prototypes}(e)) and from these LuH, LuN$_{0.5}$H$_{0.5}$, LuN$_{0.75}$H$_{0.25}$ and LuN were 0~GPa dynamically stable.  

N/H substitution into the fluorite-type, or $C1$, LuH$_2$ phase (Figure \ref{fig:prototypes}(b)), yielded another set of potential candidates (Figure \ref{fig:structures2}(f)), with LuN$_{0.5}$H$_{1.5}$ and LuNH being dynamically stable at 0~GPa. LuNH is a half-Heusler-like compound with equal amounts of N and H occupying the tetrahedral interstices of the Lu lattice. From the dynamically stable phases identified in this section, $C1$ LuN$_{0.5}$H$_{1.5}$ is weakly metallic under PBE-DFT, and thus likely in actuality to be a non-metal. The rest are metallic. Below, we will compare the pressure-volume relation calculated for the phases discussed in this section with the experimental results obtained for compounds \textbf{A} and \textbf{B}, and discuss the thermodynamic stability, electronic structure and potential for superconductivity in these prototype-based Lu-N-H phases. \\[2ex]

\noindent\textbf{Structures Inspired by Evolutionary Searches:}

\begin{figure}[b!]
   \centering
    \includegraphics[width=\columnwidth]{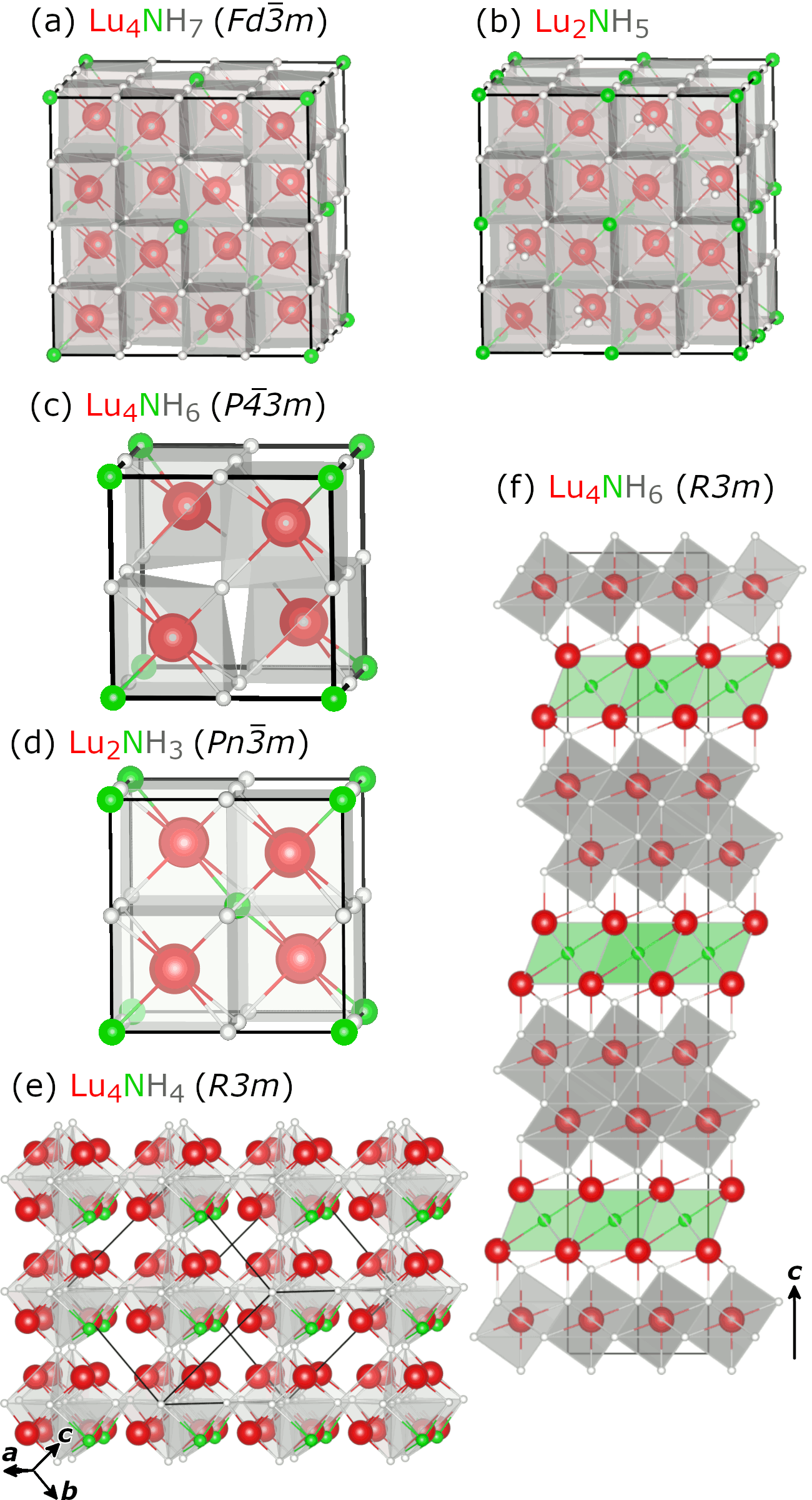}
    \caption{Crystal structures of various dynamically stable Lu-N-H phases obtained from a combination of CSP searches -- some constrained -- and subsequent modification. Lu$_2$NH$_5$ and Lu$_2$NH$_3$ had PBE band-gaps of 1.09 and 0.06~eV at 10~kbar.}
    \label{fig:structures3}
\end{figure}

Figure \ref{fig:structures3} illustrates a number of 0~GPa dynamically stable, metallic phases with \emph{fcc} Lu sublattices that were found in a variety of ways. The $Fd\bar{3}m$ Lu$_4$NH$_7$ phase (Figure \ref{fig:structures3}(a)) was found in an unconstrained evolutionary search performed at 1~GPa. It can be constructed from a modified 2$\times$2$\times$2 supercell of CaF$_2$-type LuH$_2$,  in which 1/8 of the tetrahedral interstices of the Lu lattice are occupied by N rather than H. The distribution of the N atoms throughout the unit cell is in a diamond-like lattice. In this structure, the octahedral interstices of the Lu lattice are left empty. This structure belongs to the same family of phases illustrated in Figure \ref{fig:structures2}(f), representing another N-substituted CaF$_2$-type LuH$_2$ derivative. However, rather than being derived from prototype modification, it was located in an \textsc{XtalOpt} search and then served as a template to construct additional metastable phases. One of these, $Fd\bar{3}m$ Lu$_2$NH$_5$ (Figure \ref{fig:structures3}(b)), was generated by placing H$_2$ units into some of the empty octahedral interstices of the Lu lattice, and replacing an additional H atom from Lu$_4$NH$_{7}$ by N, so that the N atoms now trace out a \emph{bcc} network within the structure, leaving H$_2$ molecules lying along only half of the N-N contacts.

Another (incomplete, or prematurely terminated) \textsc{XtalOpt} search at 1~GPa identified $P\bar{4}3m$ Lu$_4$NH$_6$ (Figure \ref{fig:structures3}(c)), which was chosen for further analysis and modification due to its dynamic stability, and the good match between its simulated XRD pattern with experiment. Like $Fd\bar{3}m$ Lu$_4$NH$_7$, $P4\bar{3}m$ Lu$_4$NH$_6$ is similarly a variant of the CaF$_2$-type LuH$_2$ structure, in which 1/8 of the tetrahedral interstices of the Lu lattice are occupied by N rather than H and an additional 1/8 of the tetrahedral interstices are left empty. Rather than the diamond-like distribution of N atoms found in Lu$_4$NH$_7$, the substituting N atoms and vacancies are arranged in a CsCl-type framework. Filling the vacancies in Lu$_4$NH$_6$ with N atoms yields the $Pn\bar{3}m$ Lu$_2$NH$_3$ structure (Figure \ref{fig:structures3}(d)).

In the above phases, N atoms were positioned in the tetrahedral interstices of an \emph{fcc} Lu framework, whereas in Lu$_2$NH$_5$ the octahedral interstices were partially occupied by H$_2$ molecular units. In $R3m$ Lu$_4$NH$_4$, which was identified using an \textsc{XtalOpt} search carried out at 6~GPa where the Lu sublattice was constrained to maintain the $Fm\bar{3}m$ space group, the N atoms are not found within the tetrahedral holes but instead lie on 1/4 of the octahedral holes of the Lu lattice (Figure \ref{fig:structures3}(e)). The N atoms in  Lu$_4$NH$_4$ trace out a simple cubic arrangement, with their positions shifted slightly off of the center of the surrounding Lu$_6$ octahedra, while the tetrahedral interstices are half occupied by H and half are left empty. The remaining H atoms can be grouped into H@H$_6$ vertex-sharing octahedra.

The unconstrained 0~GPa \textsc{XtalOpt} searches that mainly uncovered the semiconducting compounds shown in Figure~\ref{fig:structures1} also produced the metallic $R3m$ Lu$_4$NH$_6$ phase (Figure \ref{fig:structures3}(f)). In this phase, the Lu-N and Lu-H interactions become separated, with layers of N@Lu$_6$ octahedra -- in essence, slabs of $B1$ LuN -- interrupting a CaF$_2$-type packing of Lu and H. Because this phase was found using an EA search that generated sufficient structures to explore the potential energy landscape, it was 136.1~meV/atom lower in enthalpy than the previously discussed $P\bar{4}3m$ Lu$_4$NH$_6$, and at 5~GPa this difference increased to 175~meV/atom. Perhaps this structure, with N-rich layers intercalated into a LuH$_2$ matrix, could hint at a strategy for inducing epitaxial strain on simple LuH$_n$ structures, thereby altering their electronic and mechanical properties from those of their parent.

\section{Properties: Stability, Equation of States, Electronic Structure, Superconductivity}

\noindent\textbf{Thermodynamics:}

The thermodynamic stability of the new structures was investigated by calculating their formation enthalpies relative to the solid elemental phases as a function of pressure. The reference phases employed were Lu: $\alpha$-Sm (0-8~GPa \cite{Liu1975_lutetium}) and the hexagonal phase (9-10~GPa \cite{Spedding1961_lutitium}); H$_2$: $P6_3/m$ phase (0-10~GPa \cite{Pickard2007_hydrogen}); and N$_2$: $\alpha$-N$_2$ phase (0-7~GPa \cite{Donohue1961_N2}) and $\epsilon$-N$_2$ phase (8-10~GPa \cite{Mills1986_N2}). Known experimental phases including fluorite type LuH$_2$, $B1$ LuN, $P\bar{3}c1$ LuH$_3$, and $P2_13$ NH$_3$ \cite{Pickard2008_NH3} were also considered.

\begin{figure}[t!]
    \centering
    \includegraphics[width=\columnwidth]{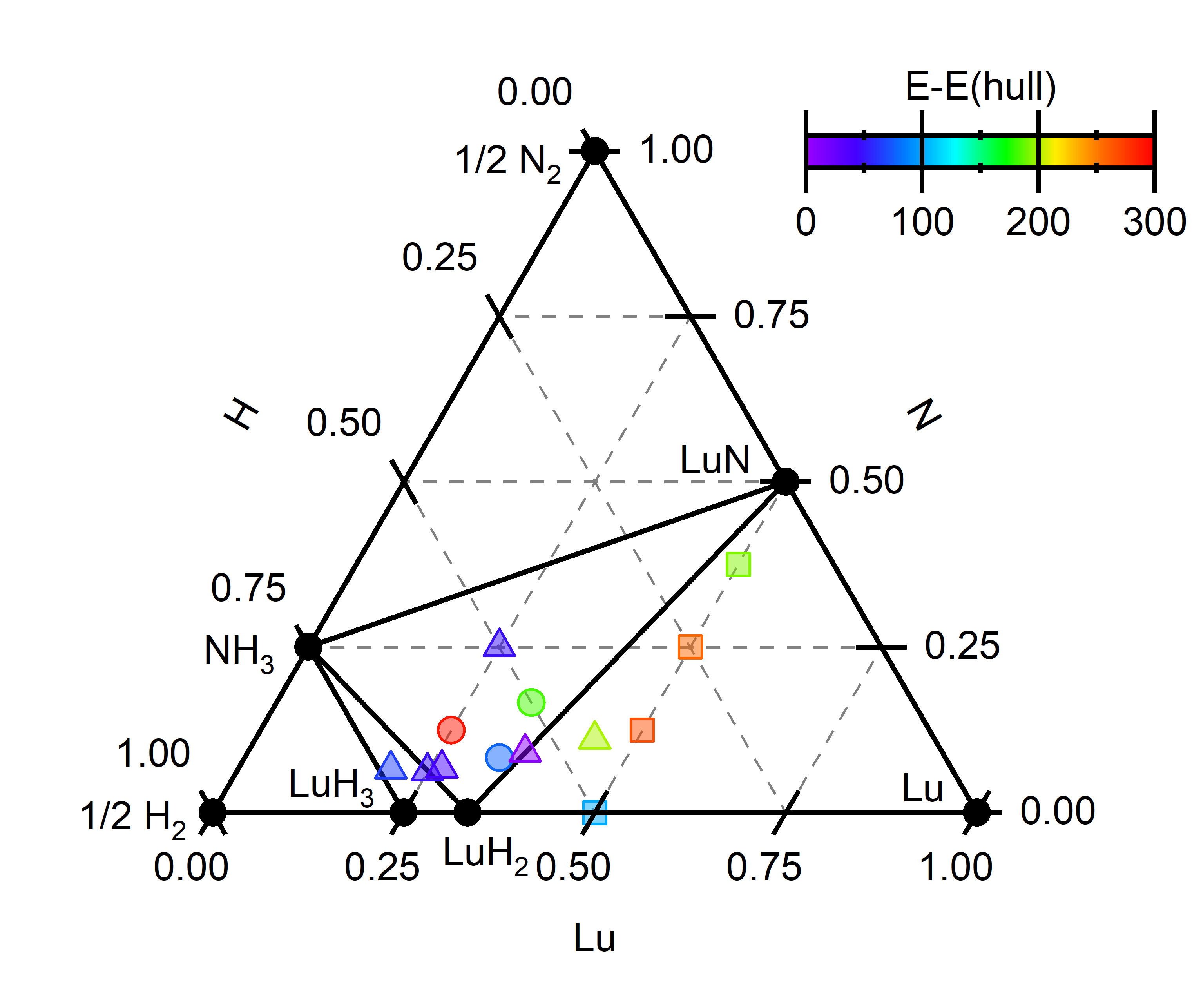}
    \caption{Convex hull at 0~GPa. Only dynamically stable structures within 300~meV/atom above the hull are shown. If multiple structures exist for the same stoichiometry, only the most stable structure is listed. Black dots represent thermodynamically stable phases on the hull, and the colored points are colored by their distance from the hull in meV/atom. Triangles: structures generated via evolutionary search. Boxes: structures from prototype modification. Circles: structures generated by inserting atoms into structures derived from EA searches.}
    \label{fig:hull_0gpa}
\end{figure}

The 0~GPa convex hull shown in Figure \ref{fig:hull_0gpa} illustrates that only the known structures are thermodynamically stable, and all of the previously discussed Lu-N-H compounds are thermodynamically unstable within the static lattice approximation. Up to 10~GPa only the known phases lie on the hull, while all others lie above it. A structure's thermodynamic stability can be characterized by its distance to the convex hull, which is listed in the Supporting Information, where it is also plotted as a function of pressure (for all compounds, regardless of their dynamic stability). % in Figure~\ref{fig:hull_eos}.  
Herein, we employ 70~meV/atom, which corresponds to the 90$^{th}$ percentile of the DFT-calculated metastability of all of the known inorganic crystalline materials~\cite{materialsproject}, as a gauge to identify those structures that could potentially be synthesized. At 0~GPa only five structures  -- all found using our unconstrained crystal structure search -- fall in this range. From these only $R3m$ Lu$_4$NH$_6$ was metallic.

Let us now turn to the metallic phases with \emph{fcc} Lu lattices.  For the rock-salt solid-solution family, hydrogen concentrations ranging from 25-100\% were roughly within 150-250~meV/atom from the hull, with $B1$ LuH as the lower boundary. For the zinc-blende solid-solution system this range expanded to 100-500~meV/atom, with $B3$ LuH corresponding to the lower boundary as well. Doping fluorite LuH$_2$ causes its energy to explode quickly: 25\% nitrogen content results in an increase of energy by $\sim$200~meV/atom above the convex hull, which rises to $\sim$550~meV/atom for a 50\% composition, and 1.1~eV/atom for 75\% nitrogen content.  The 0~GPa ternary convex hull plot shows that most of the low-enthalpy metastable structures are found at the bottom left hand corner. The reason for this is that these are the only regions where full unconstrained CSP searches were performed, and the survivor bias makes us think that this region is where stable structures might appear. It should be noted, however, that these stoichiometries were chosen because exploratory calculations suggested their volumes were likely to provide the best match with the experimental equation of states of compound \textbf{A}.  This will be explored shortly below.

Assuming linear behavior of the enthalpy-pressure relation, we were able to estimate the pressure where the considered phases may become thermodynamically stable if the slope of the distance from the hull versus pressure is negative. This estimate does not take into account the dynamic stability, nor does it include temperature or effects arising from the zero point motion of the nuclei. The results suggest that $B1$ LuN-LuH mixtures become favored at high-pressures: LuH by $\sim$30~GPa, $\sim$50~GPa for LuN$_{0.25}$H$_{0.75}$, about 70~GPa for LuN$_{0.5}$H$_{0.5}$, and 80~GPa for LuN$_{0.75}$H$_{0.25}$. The higher the hydrogen concentration, the lower the predicted stabilization pressure, with a lower boundary of 30~GPa for $B1$ LuH. The slope of the other two solid solutions considered, $B3$ and $C1$ type, is positive suggesting they will never be stabilized. Two further phases that could potentially be stabilized within the megabar range are  $R3m$ Lu$_4$NH$_6$ (16~GPa) and $P_1$ Lu$_4$NH$_{10}$ (34~GPa) because they are very close to the hull. The rest of the structures either possess a positive slope, or cannot be stabilized until at least 140~GPa. \\

\noindent\textbf{Equation of States and X-ray Diffraction Patterns:}

\begin{figure*}[!hbpt]
    \centering
    \includegraphics[width=0.95\textwidth]{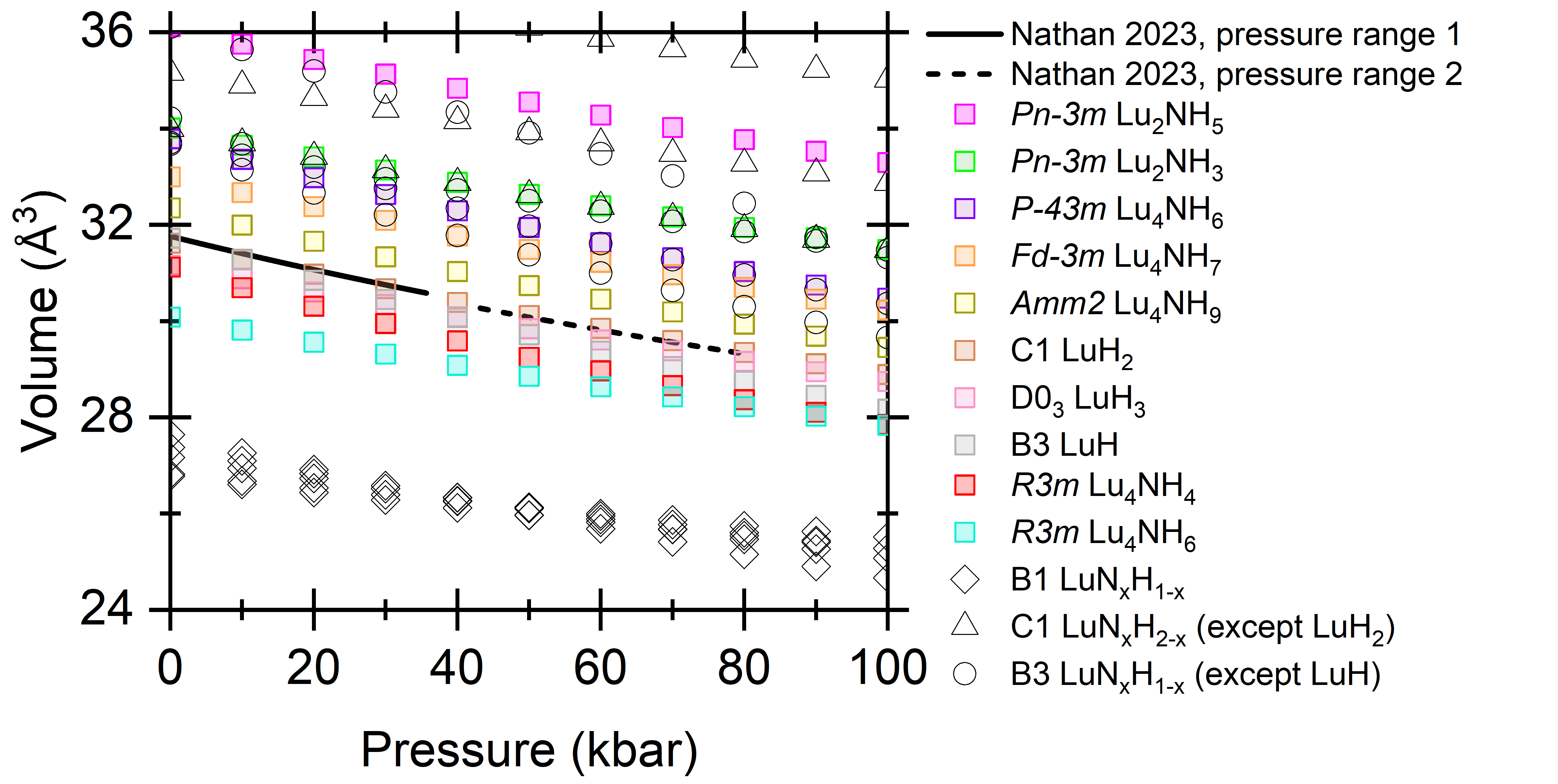}
    \caption{The DFT calculated pressure-volume relationship or equation of states (EoS) of the Lu-N-H phases considered in this study. The colored squares correspond to the specified structures and the open diamond, triangle and circles to various structures comprising the $B1$, $B3$ and $C1$ solid solution series (see Figure \ref{fig:structures2}), except for $C1$ LuH$_2$ and $B3$ LuH. The black lines represent the EoS fitted using the Birch-Murnaghan method for compound \textbf{A} using data from the pressure ranges $0<P<40$~kbar (solid) and $P>42.7$~kbar (dashed)~\cite{Dasenbrock:2023}.}
    \label{fig:volume_eos}
\end{figure*}

One of the key experimental observables guiding our choice of stoichiometries was the pressure-volume relation, or equation of states (EoS), of the majority phase presented in Reference \cite{Dasenbrock:2023}, which was assigned tentatively as an $Fm\bar{3}m$ structure with a LuH$_{3-\delta}$N$_\epsilon$ stoichiometry (or compound \textbf{A}). Above $\sim$30~kbar a first-order structural phase transition with a $\sim$0.3\% volume discontinuity was observed suggesting that the metal lattice of the resulting non-superconducting phase distorted  to the $Immm$ spacegroup. 
In Figure \ref{fig:volume_eos} we plot the EoS fits from Reference \cite{Dasenbrock:2023} for the majority phase, which were obtained for two pressure ranges. Choosing stoichiometries whose volumes matched well with experiment was initially non-intuitive because the effective radius of the metal atom changes substantially with its oxidation state being largest for Lu and smallest for Lu$^{3+}$.   

From all of the phases we considered, both fluorite LuH$_2$ and zinc-blende LuH presented the best match with the experimental data below 40~kbar. At higher pressures, however, the volume of $B3$ LuH was computed to become progressively smaller than the measured value for compound \textbf{A}. The good agreement with $C1$ LuH$_2$, on the other hand, remained up to at least 80~GPa. At 0~GPa $B3$ LuH was slightly larger than $C1$ LuH$_2$, in-line with the general notion that the effective radius of Lu$^{+}$ is larger than that of Lu$^{2+}$. However, the volume of $B3$ LuH shrinks much faster (with a slope that is similar to that of $B1$ LuH) with increasing pressure as compared to that of $C1$ LuH$_2$, while the volume of cubic LuH$_3$ shrinks at an even slower rate. Thus, the compressibility in these compounds appears to be dependent upon the repulsion exhibited between the ionic cores, with a larger number of H$^-$ anions resulting in a higher resistance to compression. Due to its larger ionic radius, N$^{3-}$ is less compressible than H$^-$. Since the computed EoS of cubic LuH$_3$ has a smaller slope than the EoS derived from experiment, and introduction of nitrogen will decrease the slope further, it could be expected that a compound with the LuH$_{3-\delta}$N$_\epsilon$ stoichiometry that was proposed for compound \textbf{A} would not have a slope that coincides with the experimentally derived EoS. \\

Because the calculated EoS of fluorite LuH$_2$ across the whole pressure range yielded the best fit with the experimentally reported EoS, we employed the quasiharmonic approximation to obtain a temperature-dependent EoS. Fitting the resulting EoS using the Birch-Murnaghan method at 300 K yielded $V_0$ (reference volume at $P=0$) of 31.85~\r{A}$^3$, $K_0$ (bulk modulus at $P=0$) of 922.8~kbar, $K_0^\prime$ ($dK_0/dP$ at $P=0$, dimensionless) of 3.7 (data for 0 and 100 K can be found in the SI). This compares well with the values presented in Ref~\cite{Dasenbrock:2023} obtained using fits to data collected below (above) 40 kbar of 31.74 (31.6)~\r{A}$^3$, 886.4(900)~kbar, and 4, respectively.

To determine if the structures discussed here could yield XRD patterns similar to those observed in experiment, their simulated 0~GPa XRD patterns were generated, as was an XRD pattern for a model $Fm\bar{3}m$ Lu cell whose lattice constant ($a=5.029$~\AA{}) was in-line with the refined unit cell suggested for superconducting compound \textbf{A} at 0~GPa (plots are provided in the Supporting Information).  The \textsc{PyXtal} XRD Similarity tool \cite{Fredericks2021_pyxtal} was used to assess the similarity between the simulated powder XRD patterns of the proposed structures and that of the model $Fm\bar{3}m$ Lu cell. The strongest matches came from the experimental phases CaF$_2$-type LuH$_2$ (0.9848), AlFe$_3$-type LuH$_3$ (0.9316), and from ZnS-type LuH (0.9962) -- in-line with the volume of $B3$ LuH adhering closely to the experimental EoS near 0~GPa. Of the N/H-doped NaCl, ZnS, and CaF$_2$-type structures, the best XRD matches could be attributed to the ZnS-based phases, with the NaCl-based phases comparing most poorly. Of the phases directly obtained from \textsc{XtalOpt} searches or based on modifications of  \textsc{XtalOpt} results, $Fd\bar{3}m$ Lu$_4$NH$_7$ and $R3m$ Lu$_4$NH$_4$ provided the best matches, although their enthalpies place them well above the convex hull in the pressure range of interest. \\

\noindent\textbf{Electronic Structure and Superconductivity:}

Superconductivity has been measured in elemental Lu at pressures above $\sim$100~kbar, with $T_c$ rising to $\sim$0.6~K near 160~kbar~\cite{Wittig1974_superconductivity}. Adding hydrogen and mild pressure does not improve the superconducting properties much or at all: superconductivity in LuH$_2$ was not observed down to 1.5~K at pressures as high as 7.7~GPa~\cite{arxiv2303.06718}. These recent experimental results are in agreement with our computed values at 10~kbar, obtained via the Allen-Dynes modified McMillan equation, which is thought to be appropriate for phonon-mediated superconductors whose $\lambda < \sim$1-1.5. As shown in Table \ref{tab:tc}, we found that the $T_c$ of fluorite-type LuH$_2$ was $\sim$0.1~K, owing to a small $\omega_\text{ln}$ combined with a modest $\lambda=0.29$. %This result was obtained using a $\mu^*=0.1$, and larger values will yield even smaller $T_c$s.  

To study the potential for superconductivity in ternary Lu-N-H compounds we performed EPC calculations for the previously discussed metallic phases that were dynamically stable at 10~kbar -- the pressure at which the maximum $T_c$ was observed in Reference \cite{Dasenbrock:2023}.  Table \ref{tab:tc} shows that though the $T_c$s of most of these phases (with the exception of LuN$_{0.5}$H$_{1.5}$) were predicted to surpass that of $C1$ LuH$_2$, they do not even reach the boiling point of liquid nitrogen, in agreement with recent theoretical calculations that did not find any Lu-N-H phases with room temperature superconductivity~\cite{arxiv2303.12575}. 
\begin{table}[!hbpt]
    \centering
    \begin{tabular}{l|c|c|c}
    \hline 
    Structure     &   $\lambda$   &   $\omega\textsubscript{ln}$ (K)  &
$T\textsubscript{c}$ (K) \\
    \hline 
   CaF$_2$-type LuH$_2$       &   0.29     &   302 &   0.09               \\
   CaF$_2$-type LuNH          &   0.78     &   377  &   16.9 (18.3)                \\
   CaF$_2$-type LuN$_{0.5}$H$_{1.5}$   & 0.11 & 680   & 0.0  \\
   $R3m$ Lu$_4$NH$_4$         &   0.64     &   151    & 4.2  \\
   $P\bar{4}3m$ Lu$_4$NH$_6$  &   0.48     &   291    & 2.9  \\
   $Fd\bar{3}m$ Lu$_4$NH$_7$ & 0.47 & 435 & 4.2 \\
 $R3m$ Lu$_4$NH$_6$ & 0.29 & 306 & 0.12 \\
   \hline
    \end{tabular}
    \caption{The electron phonon coupling, $\lambda$, logarithmic average frequency, $\omega\textsubscript{ln}$, and superconducting critical temperature, $T_c$, estimated using the Allen-Dynes modified McMillan equation with $\mu^*=0.1$ at 10~kbar for select Lu-N-H compounds. For LuNH the numerical solution of the Eliashberg equations was employed to obtain the value in brackets.}
    \label{tab:tc}
\end{table}

The highest $T_c$ compound we found, fluorite type LuNH, can be derived from LuH$_2$ by replacing 50\% of the hydrogen atoms by nitrogen (Figure \ref{fig:structures2}(f)). This chemical substitution dramatically increased the EPC, placing it in the realm of the ambient pressure conventional superconductor with the highest confirmed $T_c$, MgB$_2$. However, the larger $\lambda$ of 0.78 was attained at a cost of the thermodynamic stability: while $C1$ LuH$_2$ fell on the 10~kbar hull, LuNH was 564~meV/atom above the hull, suggesting it could never be made. The $T_c$ of LuNH ($\sim$17~K) was estimated to be a factor of two smaller than that of MgB$_2$ with its strong covalent B-B bonds, whose motions, with frequencies around 600~cm$^{-1}$, yield an $\omega\textsubscript{ln}$ of 504~cm$^{-1}$ (or 725~K) \cite{kong2001}. As we shall soon see, in LuNH the EPC are relatively evenly distributed from the high frequency motions of the hydrogen vibrations to the very low frequency acoustic modes. Their $\alpha^2F$-weighted logarithmic average yields an $\omega\textsubscript{ln}$ of 257.2~cm$^{-1}$ (or $\sim$370~K). Numerical solution of the Eliashberg equations raised the $T_c$ of LuNH only slightly to $\sim$18~K. 

\begin{figure}[!hbpt]
    \centering
    \includegraphics[width=\columnwidth]{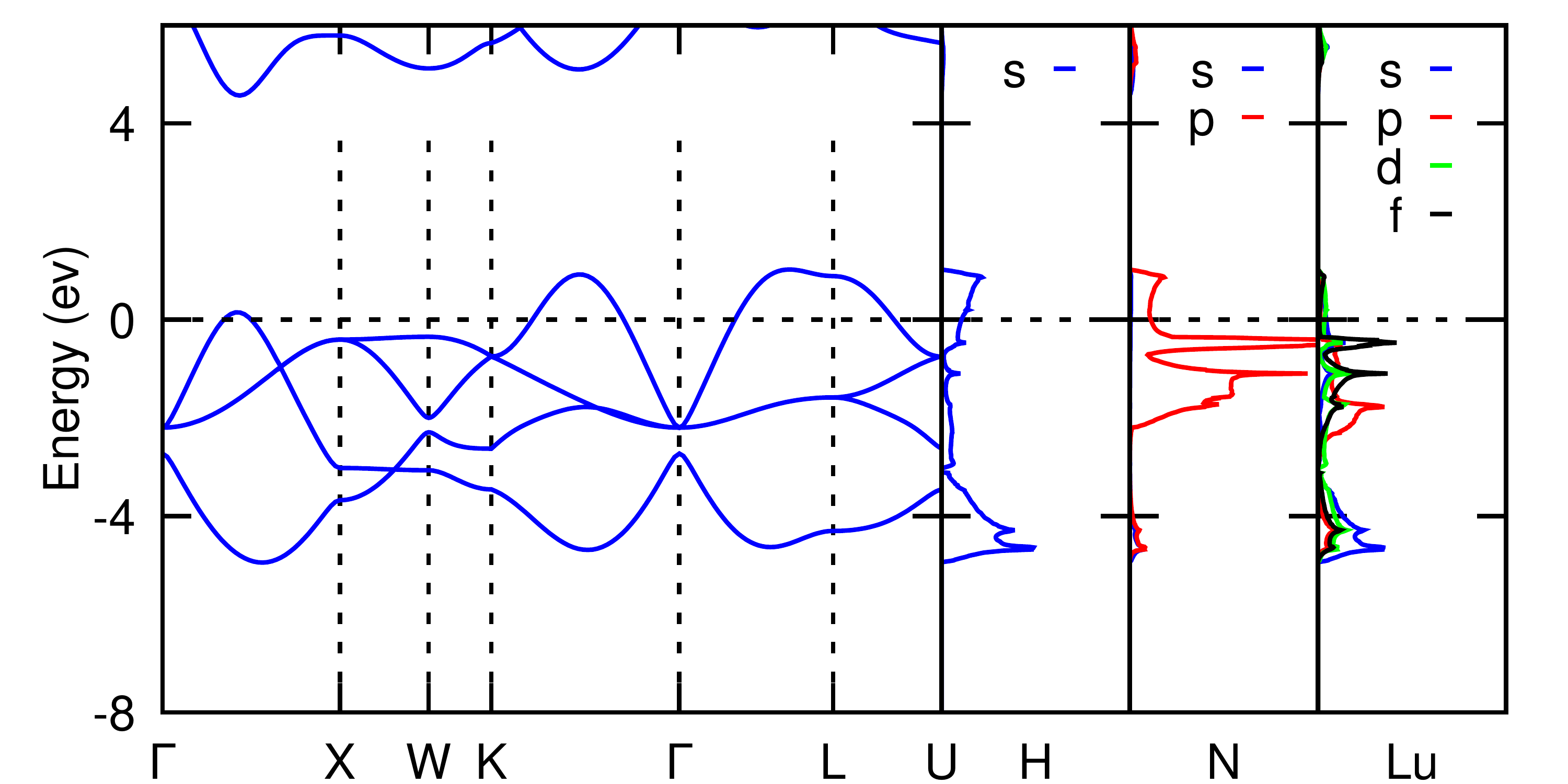}
    \caption{PBE band structure and projected densities of states of fluorite-type LuNH at 10~kbar.}
    \label{fig:band}
\end{figure}

Let us examine the electronic structure of CaF$_2$-type LuNH and its contributions to the EPC to better understand how these factors influence the $T_c$. Replacing H by N in LuH$_2$ increases the density of states (DOS) at the Fermi level ($E_F$) by around 50\% from 0.019~states/eV/\r{A}$^3$ to 0.027~states/eV/\r{A}$^3$, concomitantly increasing the $T_c$. As shown in Figure \ref{fig:band}, the major contributions to the DOS at $E_F$ are the H $1s$ and N $2p$ states, with a negligible amount from the metal, indicative of a +3 oxidation state. The primitive cell of LuNH contains one formula unit, and as a result its conduction band is half-filled. The reaction $\text{LuNH}+\frac{1}{2}\text{H}_2 \rightarrow \text{LuN}+\text{H}_2$ is exothermic by 400~meV/atom; we would therefore expect the products of this reaction to be found in a CSP search for unit cells whose sizes approach infinity.

\begin{figure}[!hbpt]
    \centering
    \includegraphics[width=\columnwidth]{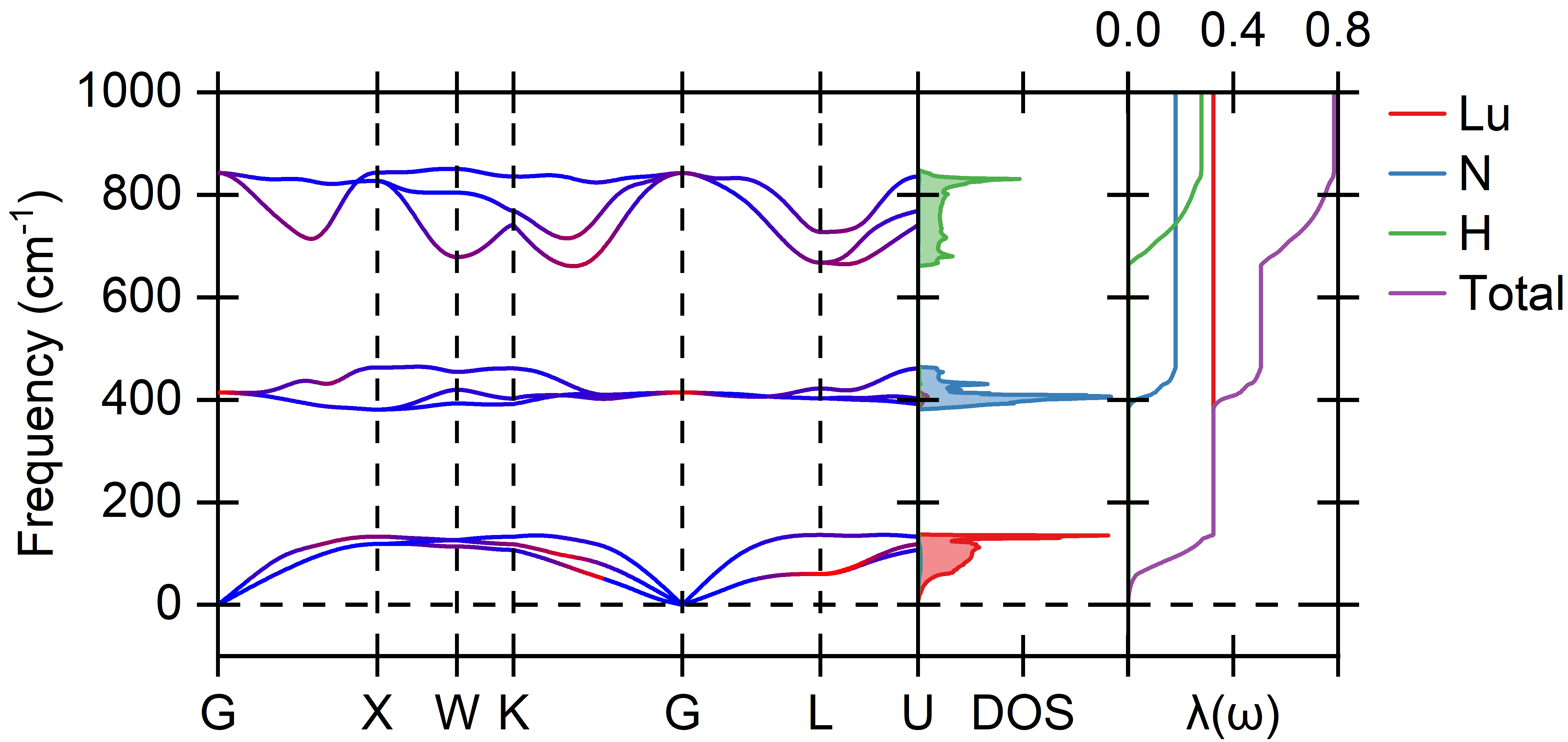}
    \caption{Phonon dispersion curve and projected EPC constant ($\lambda_{\mathbf{q}\nu}$). Blue color indicates $\lambda_{\mathbf{q}\nu}$ approaches 0, and red indicates $\lambda_{\mathbf{q}\nu}$ approaches the maximum value of 0.36. The atom projected phonon density of states is illustrated, along with the total $\lambda$ and the integral of $\lambda(\omega))$ separated into regions comprising the Lu, N and H-based modes.}
    \label{fig:epc_lunh}
\end{figure}

Pivoting to the phonon band structure in Figure \ref{fig:epc_lunh}, we observe that the large differences in the mass between the three elements splits their bands nicely into three regions. The vibrational modes of lutetium are mainly below 140~cm$^{-1}$ (acoustic region), nitrogen are between 380-470~cm$^{-1}$ and hydrogen above 660~cm$^{-1}$. It should be noted that due to the extremely heavy mass of lutetium versus nitrogen and hydrogen (175 vs.\ 14 and 1~a.u.), lutetium moves roughly ten times slower than the hydrogen, and four times slower than the nitrogen. As a result, the atomic displacements of the nitrogen and hydrogen atoms along the low frequency acoustic modes are still significant. 

Because of the separation of these vibrational modes, their contribution to the total EPC can be obtained: motions from the acoustic modes contribute 41\%, 23\% from the nitrogen active region, and 35\% from the hydrogen active region. The largest Lu-based contribution originates from the lower two acoustic phonon branches around the middle of the $\Gamma$-$K$ path, and also around the $L$ point. Visualization of these motions show they result in the formation of N-Lu-H molecular fragments and a hexagonal-like Lu lattice. In the nitrogen-active region the largest EPC is found at the $\Gamma$ point, resulting from the nitrogen atoms approaching lutetium to form N-H motifs.  In the hydrogen-active region, the entire bands exhibit moderate EPC, especially at several points where the modes are softened; visualization shows that these correspond to the motion of hydrogen atoms closer to lutetium to form H-Lu units.

The $\omega\textsubscript{ln}$ of our Lu-N-H compounds ranged from $\sim$150~K ($R3m$ Lu$_4$NH$_4$) to 680~K (CaF$_2$-type LuN$_{0.5}$H$_{1.5}$). The absence of high frequency vibrations in these compounds, resulting from the low pressure and the absence of covalent bonds, suggests that higher $\omega\textsubscript{ln}$ are unlikely to be found in other Lu-N-H compounds at 10~kbar with \emph{fcc} Lu lattices. Generally speaking, the $\omega\textsubscript{ln}$ calculated for hydrogen and the high-$T_c$ hydrides at extreme pressures is significantly higher, with values of 1200-1800~K not being uncommon. For those hydrides where comparable $\omega\textsubscript{ln}$ values have been calculated, room temperature superconductivity has only been predicted in phases with a very large EPC  (e.g.\ $Fmmm$ ThH$_{18}$ at 400~GPa, $\omega\textsubscript{ln}=$568~K, $\lambda=$3.39, $T_c=$296~K)~\cite{Zhong2022_JACS}. Therefore, we speculate that similar EPC constants are required for a Lu-N-H compound to be superconducting near room temperature, provided the mechanism is phonon-mediated.

\section{Conclusions} 

Density functional theory calculations were performed to explore Lu-N-H containing compounds that could be (meta)stable in a pressure range of about 0-100~kbar (10~GPa). The computations were biased towards systems where the Lu atoms adopt an \emph{fcc} arrangement, because it was recently suggested that a compound with this structural feature could be responsible for the near-ambient superconducting critical temperature, $T_c$, of 294~K reported at 10~kbar~\cite{Dasenbrock:2023}. Based on the results of our calculations we conclude that:
\begin{itemize}
 \item The Lu-N-H potential energy landscape, within the static lattice approximation and neglecting quantum nuclear and anharmonic effects, contains many local minima with \emph{fcc} Lu lattices. Other geometries, not explicitly considered here, could be generated via altering the N/H ratio of the solid-solution prototypes we discuss. Which of these structures are synthesizable, and which are kinetically and/or thermally stable and relatively chemically inert is currently unknown.
 \item None of the ternary compounds studied here are thermodynamically stable (e.g.\ they do not lie on the convex hull) up to 10~GPa at 0~K. Only the known binaries, LuH$_2$, LuH$_3$, LuN and NH$_3$, comprise the convex hull. Thermal effects and the role of configurational entropy on the thermodynamic stability are not known.
 \item From all of the phases considered here, the one whose equation of states (EoS) had the closest match with the fits to experimental data obtained for compound \textbf{A} was fluorite-type LuH$_2$, with errors smaller than 0.3\% up to 80~kbar. EoS calculations on model compounds suggest that the previously proposed formula for compound \textbf{A}, LuH$_{3-\delta}$N$_\epsilon$, would not have the same slope as what was observed experimentally. 
 \item XRD similarity indices for the compounds studied here compared to a pure \emph{fcc} Lu lattice with the experimental lattice constant indicated a fair match at 0~GPa for binaries LuH$_2$, LuH$_3$, and ZnS-type LuH, N-substituted ZnS-type LuH, $Fd\bar{3}m$ Lu$_4$NH$_7$ and $R3m$ Lu$_4$NH$_4$. 
 \item Many, though not all, of the investigated phases exhibited metallic behavior, and their density of states at the Fermi level (DOS at $E_\text{F}$) varied greatly.  For example, the main contributions to the DOS at $E_\text{F}$ for $B1$ and $B3$ LuH were the Lu $d$ states; for LuH$_2$ the DOS at $E_\text{F}$ was very small and mainly lutetium $p$-like, and for LuNH the main components arose from hydrogen $s$ and nitrogen $p$ states. 
 \item The logarithmic average frequency, $\omega_\text{ln}$, of the Lu-N-H compounds whose superconducting properties we studied ranged from $\sim$150-680~K, and the electron phonon coupling constants, $\lambda$, varied between 0.1-0.8. Assuming conventional superconductivity, the $T_c$s of such compounds can be estimated using the modified McMillan Allen Dynes equation. Under this approximation, and with $\mu^*=0.1$ at 10~kbar we obtain a $T_c$ of 0.09~K for fluorite LuH$_2$. The highest $T_c$ compound we found was fluorite-type LuNH with a $T_c$ of 17~K. 
\end{itemize}

Though we have not uncovered an Lu-N-H containing phase with a superconducting critical temperature near what was recently reported in Reference \cite{Dasenbrock:2023}, we believe our computations shed light on the structures that contain these elemental combinations at mild pressures. Future work will ascertain if our choice of standard DFT parameters (gradient corrected exchange functional, neglect of spin polarization and strong electron correlations, and inclusion of $f$ electrons in the core) impact our conclusions. Our work also highlights the complexity inherent in the computational search for phases that may be metastable with desired structural and property characteristics in multi-element \emph{ab initio} (or even machine-learning-assisted) crystal structure prediction. 

\section{Acknowledgments}
We are grateful to R.\ Dias for sharing experimental data, as well as G.W.\ Collins and R.J.\ Hemley for useful discussions.
K.H.\ acknowledges the Chicago/DOE Alliance Center under Cooperative Agreement Grant No.\ DE-NA0003975, and N.G.\ the U.S. National Science Foundation (DMR-2132491). This material is based upon work supported by the U.S. Department of Energy, Office of Science, Fusion Energy Sciences funding the award entitled High Energy Density Quantum Matter under Award Number DE-SC0020340. Computations were carried out at the Center for Computational Research at the University at Buffalo (http://hdl.handle.net/10477/79221).

\bibliographystyle{apsrev4-1}
\bibliography{ClathrateCP,LuNH}

\end{document}